# Hybrid, Gate-Tunable, van der Waals p-n Heterojunctions from Pentacene and MoS$_2$

*Deep Jariwala[1†], Sarah L. Howell[1†], Kan-Sheng Chen[1], Junmo Kang[1], Vinod K. Sangwan[1], Stephen A. Filippone[1], Riccardo Turrisi[2], Tobin J. Marks[1,2]\*, Lincoln J. Lauhon[1]\* and Mark C. Hersam[1,2]\**

[1]Department of Materials Science and Engineering, Northwestern University, Evanston, Illinois 60208, USA.

[2]Department of Chemistry, Northwestern University, Evanston, Illinois 60208, USA.

† These authors contributed equally.

*e-mail: m-hersam@northwestern.edu, lauhon@northwestern.edu, t-marks@northwestern.edu

**ABSTRACT**

The recent emergence of a wide variety of two-dimensional (2D) materials has created new opportunities for device concepts and applications. In particular, the availability of semiconducting transition metal dichalcogenides, in addition to semi-metallic graphene and insulating boron nitride, has enabled the fabrication of 'all 2D' van der Waals heterostructure devices. Furthermore, the concept of van der Waals heterostructures has the potential to be significantly broadened beyond layered solids. For example, molecular and polymeric organic solids, whose surface atoms possess saturated bonds, are also known to interact *via* van der Waals forces and thus offer an alternative for scalable integration with 2D materials. Here, we demonstrate the integration of an organic small molecule p-type semiconductor, pentacene, with a 2D n-type semiconductor, MoS$_2$. The resulting p-n heterojunction is gate-tunable and shows asymmetric control over the anti-ambipolar transfer characteristic. In addition, the pentacene/MoS$_2$ heterojunction exhibits a photovoltaic effect attributable to type II band alignment, which suggests that MoS$_2$ can function as an acceptor in hybrid solar cells.

TOC GRAPHIC:

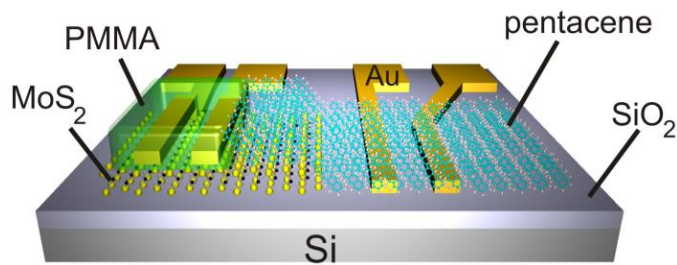

**TOC Figure**

TEXT: The isolation of layered materials as passivated, dangling bond-free monolayers presents a unique platform for integration in previously unexplored device structures.[1, 2] For instance, semi-metallic graphene, insulating boron nitride, and semiconducting $MoS_2$ have been stacked together to yield atomically thin memory devices, photodetectors, and field-effect transistors.[3-6] This principle has also been exploited to fabricate heterojunctions between conventional silicon and 2D materials[7-9] as well as carbon nanotubes with 2D materials[10] and amorphous oxide semiconductors.[11] Similarly, organic molecular and polymer semiconductors are free of dangling bonds and native surface oxides, which suggests new opportunities for van der Waals heterostructures. Despite this potential, the integration of organic semiconductors with 2D materials has thus far been limited primarily to templating ordered film growth to improve the performance of the channel or contacts in conventional field-effect transistor (FET) geometries.[12-16] Due to their relatively large light capture cross-sections, organic semiconductors also hold promise for enhancing the performance of photovoltaics based on 2D materials. Although some preliminary studies have appeared in this regard,[17, 18] a clear understanding of charge transport across the interface between organic and 2D semiconductors is lacking.

In an effort to better understand and exploit the potential of van der Waals heterostructures between organic and 2D semiconductors, we explore here the electronic and optoelectronic response in p-n heterojunctions based on pentacene and $MoS_2$. The operating principles and band profiles of this system are characterized by direct charge transport measurements, scanning photocurrent microscopy, electrostatic force microscopy, and finite element modeling. This comprehensive experimental and computational study reveals that pentacene forms a type-II heterojunction with $MoS_2$, which yields asymmetric anti-ambipolar transfer curves when operated as a three-terminal, gate-tunable diode. Furthermore, these p-n heterojunctions possess a clear photovoltaic response upon optical irradiation, which suggests that $MoS_2$ can function as an acceptor in hybrid solar cells.

Devices were fabricated from mechanically exfoliated $MoS_2$ flakes. Electron beam lithography was used to define the electrodes and junction area. A pentacene film was then deposited via thermal evaporation. We have fabricated > 10 devices in total and all of them show qualitatively similar behavior. Results from two representative devices (2L $MoS_2$/40nm pentacene) are shown in the figures. The resulting four-electrode test structure consists of a $MoS_2$ FET, pentacene/$MoS_2$ heterojunction, and a pentacene FET in series (representative device in Fig. 1a, from left to right). The heterojunction region lies within the rectangular opening in the PMMA film (Fig. 1b), which defines a partially overlapping region of a $MoS_2$ flake and the pentacene film (Fig. 1a-b). This geometry of partially overlapping semiconductors allows direct and separate probing of the contact and heterojunction interfaces with scanning probe techniques while simultaneously applying a gate voltage, which is not possible in a truly vertical device geometry.

In atomic force microscopy (AFM) topography imaging (Fig. 1c), the abrupt change in the pentacene film grain size coincides with the boundary with the underlying $MoS_2$. Because pentacene and $MoS_2$ have complementary p-type and n-type doping with electron affinities of 2.5 eV[19] and 4.2 eV (bilayer)[20, 21], respectively, and optical band gaps $\geq$ 1.9 eV, the band alignment at the pentacene/$MoS_2$ heterojunction is expected to be type II as shown in Fig. 1d.

The type II alignment results in a built-in potential and rectifying $I_D$-$V_D$ characteristics. Fig. 2a shows a three-dimensional surface plot of the $I_D$-$V_D$ (output) characteristics at different $V_G$ values. For the pentacene/$MoS_2$ heterojunction, $V_D$ refers to the bias on the pentacene electrode (electrode 3 in Fig. 1b) such that $V_D > 0$ corresponds to forward bias, while the $MoS_2$ electrode (electrode 2 in Fig. 1b) is grounded. The heterojunction transitions from a nearly insulating state at either extremes of the $V_G$ range ($V_G$ = 60 V and $V_G$ = -80 V) to a highly rectifying output behavior at intermediate values. The transfer plots in Fig. 2b further demonstrate the gate-tunability of the current through the p-n heterojunction. The transfer curve of the heterojunction (green) shows an anti-ambipolar response, which has also been observed in 1D-2D CNT-$MoS_2$,[10] 2D-2D $MoS_2$-$WSe_2$[22, 23] and 1D-3D CNT-IGZO[11] p-n heterojunctions. In contrast to these earlier devices in which the transfer characteristics are symmetric, the pentacene/$MoS_2$ heterojunction transfer characteristics are asymmetric with different slopes (*i.e.*, transconductances) on either side of the $I_D$ peak (see Supporting Information Section S2 for more details). Asymmetry in the anti-ambipolar transfer characteristic presents potential advantages in emerging circuit applications[11] and could be harnessed to achieve simultaneous phase[11] and amplitude[24] shift keying for wireless telecommunication technologies.

To identify the origin of the observed transconductance asymmetry, finite element simulations were performed (using Sentaurus TCAD) that numerically solve the Poisson and continuity equations governing free carrier transport. Where available, materials parameters were taken from the literature.[20, 21] The remaining parameters (*e.g.*, mobility and trap concentrations) used in the heterojunction device model (Fig. 2d) were informed by first modeling the unipolar FETs (Fig. 2c) on either side of the p-n heterojunction (see Supporting Information Section S4, Table S1, S2 for more details). Our model assumes the pentacene material assembled on $MoS_2$ has the same material parameters (*e.g.*, doping, mobility, etc.) as that assembled on $SiO_2$, although minor changes in substrate roughness and type can influence the molecular assembly of the overlying pentacene.[16, 25, 26] We find that increasing (decreasing) the ratio of the $MoS_2$ to pentacene mobility (Fig. 2e) or channel length (Fig. 2f) leads to a left (right) and right (left) leaning anti-ambipolar asymmetry, respectively. In other words, the anti-ambipolar asymmetry can be controlled by the ratio of the series resistances of the two semiconductor channels. The electrostatic force microscopy (EFM) measurements discussed below support this interpretation that the dominating resistances of the channels govern the current flow in forward bias and thus influence the asymmetry of the anti-ambipolar response.

The type II alignment of the pentacene/$MoS_2$ heterojunction suggests that $MoS_2$ can act as an acceptor in a photovoltaic cell. Fig. 3 shows $I_D$-$V_D$ curves in the dark and when the pentacene/$MoS_2$ heterojunction is illuminated. As expected for a type II heterojunction, a photovoltaic effect is observed upon illumination with an open circuit voltage $V_{OC}$ ~ 0.3 V and a short circuit current -$I_{SC}$ ~ 3 nA. We observe that $I_{SC}$ steadily increases with decreasing gate bias while the $V_{OC}$ remains nearly constant as a function of $V_G$ (Fig. 3b-c). The relative insensitivity of $V_{OC}$ to $V_G$ has been observed previously in $MoS_2$-$WSe_2$ junctions[22] and is attributed to the shared gate causing minimal net separation between the Fermi levels on either side of the heterojunction. The spectral response of $I_{SC}$ exhibits a peak between excitation wavelengths of 600 nm and 650 nm (Fig. 3d), which is consistent with the optical absorption profile of pentacene. Since both pentacene (see absorption spectrum in Supporting Information Section S1) and MoS$_2$ absorb in the same range and have similar absorption coefficients,[27, 28] the thicker pentacene layer is estimated to absorb at least 20 times more light than MoS$_2$ in these devices. The power conversion efficiencies are rather low (~0.004% at 625 nm) as a result of the suboptimal photocurrent collection efficiency of the lateral device geometry.

Scanning photocurrent microscopy (SPCM) and modeling are next employed to further explore the photovoltaic performance of the pentacene/MoS$_2$ heterojunction. During SPCM, a diffraction-limited laser beam is scanned over the device while the current or voltage is recorded as a function of position. Fig. 4a shows an optical micrograph of a representative device where the yellow outlined box highlights the area that was interrogated with SPCM. The dashed white line indicates the MoS$_2$ flake boundary. Fig. 4b shows the corresponding map of photocurrent acquired at $V_D = V_G = 0$ V, and Fig. 4c contains the line profiles of the photocurrent at various drain biases. The fact that the photocurrent maximum traces the MoS$_2$ flake boundary in Fig. 4b indicates that the photovoltage measured in Fig. 3 arises from the pentacene/MoS$_2$ junction and not from built-in fields at the Schottky contacts. In addition, the photocurrent originating from the metal-pentacene junction, which becomes significant at positive drain biases (Fig. 4c), is of the opposite sign and therefore counteracts the photocurrent from the junction. Therefore, the Voc measured in Fig. 3 is likely reduced by the close proximity of the pentacene/MoS$_2$ junction to the non-Ohmic contact. Fig. 4d shows a schematic of the simulated device geometry. Band-bending, which indicates the presence of electric fields that separate photogenerated charge carriers, is observed in the simulated band diagrams ($V_D = 0$ V) in both the vertical and horizontal directions (Figs. 4e-f). The flat band condition occurs near $V_D = 0.6$ V (Fig. 4g), in agreement with our experimental observation of zero photocurrent at the heterojunction near that biasing condition (Fig. 4c, green curve), indicating that the Voc could reach 0.6 V for ideal contacts.

Furthermore, the SPCM results reveal that $I_{SC}$ from pentacene/MoS$_2$ heterojunctions could also be substantially improved by modifying the geometry to increase the carrier collection efficiency. While carriers generated near the edge of the MoS$_2$ flake are collected in the current device geometry, there is minimal photocurrent signal generated by the rest of the overlapped pentacene/MoS$_2$ heterojunction area for two reasons. First, excess carriers generated far from the edge of the pentacene/MoS$_2$ heterojunction area only experience a vertical electric field, and must diffuse towards the contacts, while carriers generated at the edge of the flake experience a lateral electric field that can drive them towards the contacts. Second, carriers generated in the overlapped pentacene/MoS$_2$ region likely have a much lower minority carrier diffusion length, $L_D = (\mu \cdot \tau \cdot k_B T)^{0.5}$ where $\mu$ is the carrier mobility, $\tau$ is the carrier lifetime, $k_B$ is the Boltzmann constant, and T is the temperature. As pointed out earlier in Fig. 1d, the grain size of the pentacene is smaller on MoS$_2$ than on the surrounding SiO$_2$, which increases charge carrier scattering and trapping, resulting in lower values for $\mu$ and $\tau$.[29-31] A small $L_D$ combined with a lateral geometry for charge collection thus minimizes the collection of charge carriers from the overlapped region. Consistent with organic photovoltaics, these results suggest that a vertical bulk heterojunction geometry would allow for improved pentacene/MoS$_2$ photovoltaic device metrics.

While photocurrent is not observed throughout the overlapped pentacene/MoS$_2$ region, the above discussion and modeling indicates that a vertical built-in field is still present. To confirm the existence of this vertical electric field, we performed surface potential measurements using electrostatic force microscopy (EFM). EFM records both the sample topography and the phase shift of the cantilever vibration resulting from electrostatic force gradients (Fig. 5a,b; Supporting

Information Fig. S3). The shifted phase angle color maps shown in Figs. 5c and e are proportional to $(V_{tip}-V_{sample})^2$, where $V_{tip}$ is the tip bias and $V_{sample}$ is the sample surface potential, which varies with electrode biasing conditions. Under a large reverse bias (Fig 5c), there is a sharp potential drop at the end of the $MoS_2$ flake, and the surface potential is flat in other parts of the device channel, in agreement with the corresponding simulated band profile in Fig. 5d. At large forward bias (Fig. 5e), the junction is no longer the most resistive part of the circuit. Rather, the pentacene on $SiO_2$ acts as a current-limiting series resistor, resulting in a gradual drop in potential across the pentacene channel, again corroborated by the corresponding simulated profile (Fig. 5f). The above surface potential profiles reveal that the underlying $MoS_2$ depletes the entire 2D junction area under reverse bias. While lateral charge transport is required in these devices based on the position of the electrodes, EFM provides direct evidence of band bending in the vertical direction over the entire overlapped pentacene/$MoS_2$ junction region, which could be exploited for carrier separation in vertically oriented photovoltaics.

In conclusion, we have fabricated and characterized an organic/2D semiconductor van der Waals p-n heterojunction using pentacene and $MoS_2$. The atomically thin $MoS_2$ allows the junction to be gate-tunable with an asymmetric anti-ambipolar transfer characteristic. Spatial photocurrent mapping and surface potential profiles combined with finite element simulations confirm band bending in both the lateral and vertical directions and a type-II band alignment. The observation of a photovoltaic effect implies that $MoS_2$ could be an alternative non-fullerene acceptor for organic photovoltaics and other hybrid bulk heterojunction solar cells. Considering that optimized single-junction pentacene/$C_{60}$ cells reported in the literature have $V_{OC} \leq 0.4$ V,[32, 33] the results presented here are encouraging, especially considering the lateral device geometry and absence of any interfacial layers. While this study demonstrates the concept of an organic-2D $MoS_2$ hybrid photovoltaic device, semiconducting polymers could offer higher performance in large-area devices. With the availability of a library of organic molecules and polymers[34, 35] and solution processed[36] as well as chemical vapor deposited[37] ultrathin transition metal dichalcogenides, the concepts presented here can likely be scaled up to achieve high-performance organic/2D semiconductor van der Waals heterojunctions for use in gate-tunable diodes, hybrid photovoltaic devices, and related electronic and optoelectronic technologies.

**METHODS**

**Materials synthesis and deposition**. $MoS_2$ flakes were deposited on 300 nm $SiO_2$/Si substrates via mechanical exfoliation of bulk crystals (SPI supplies). The resulting $MoS_2$ flakes were identified by optical microscopy and later characterized by Raman spectroscopy to determine the flake thickness (Supporting Information Section S1). Pentacene powder was purchased from Sigma Aldrich and purified further by multiple sublimation/deposition cycles under reduced pressure in a thermal gradient sublimation module (Supporting Information Section S1). Next, 50 mg of the purified pentacene powder was loaded into a tungsten boat and placed in a thermal evaporator. The pentacene deposition was carried out in vacuum at a pressure $< 5.0 \times 10^{-5}$ Torr at a rate of 0.1-0.2 Å/sec. The substrate holder was not rotated and was maintained at room temperature throughout the deposition.

**Device fabrication and electrical measurements.** Devices were fabricated using standard electron beam lithography (EBL) and thermal metal evaporation. In the second step of lithography, a window was opened atop the junction area using EBL, and a shadow mask (1 mm x 1 mm square) was used to limit the evaporated pentacene area. All electrical measurements were performed under high vacuum ($P < 5 \times 10^{-5}$ Torr) in a Lake Shore CRX 4K probe station using Keithley 2400 source meters and custom LabView programs (see Supporting Information Sections S1 and S2 for more details).

**Simulations.** Sentaurus TCAD was used to model 2D devices by numerically solving the steady-state Poisson and continuity equations that govern free carrier transport using the finite element method. When available, the materials parameters in the simulations were taken from the literature. Other parameters were constrained by experimental measurements on uniform control FETs. Transport through the semiconductor junction was modeled by thermionic emission, and the contacts were modeled as Schottky barriers. Further simulation details can found in Supporting Information Section S4.

## ASSOCIATED CONTENT
**Supporting Information**:
Additional details on fabrication, electrical characterization, and simulations accompanies this paper and is available free of charge via the Internet at http://pubs.acs.org.

## AUTHOR INFORMATION
**Corresponding authors:**
*E-mail: m-hersam@northwestern.edu, lauhon@northwestern.edu, t-marks@northwestern.edu
† Deep Jariwala and Sarah L. Howell contributed equally.

## NOTES:
**Competing financial interests**: The authors declare no competing financial interests.

## ACKNOWLEDGEMENTS
This research was supported by the Materials Research Science and Engineering Center (MRSEC) of Northwestern University (NSF DMR-1121262), and the 2-DARE program (NSF EFRI-143510). D.J. acknowledges additional support from an SPIE education scholarship and IEEE DEIS fellowship. This work made use of the Northwestern University Atomic and Nanoscale

Characterization Experimental Center (NUANCE), which has received support from the NSF MRSEC (DMR-1121262), State of Illinois, and Northwestern University. In addition, the Raman instrumentation was funded by the Argonne-Northwestern Solar Energy Research (ANSER) Energy Frontier Research Center (DOE DE-SC0001059). The authors thank A. Walker, K. Stallings, and S. Clark for advice and assistance with pentacene deposition processes as well as J. Kang for help with the optical absorbance measurements.

**FIGURES**

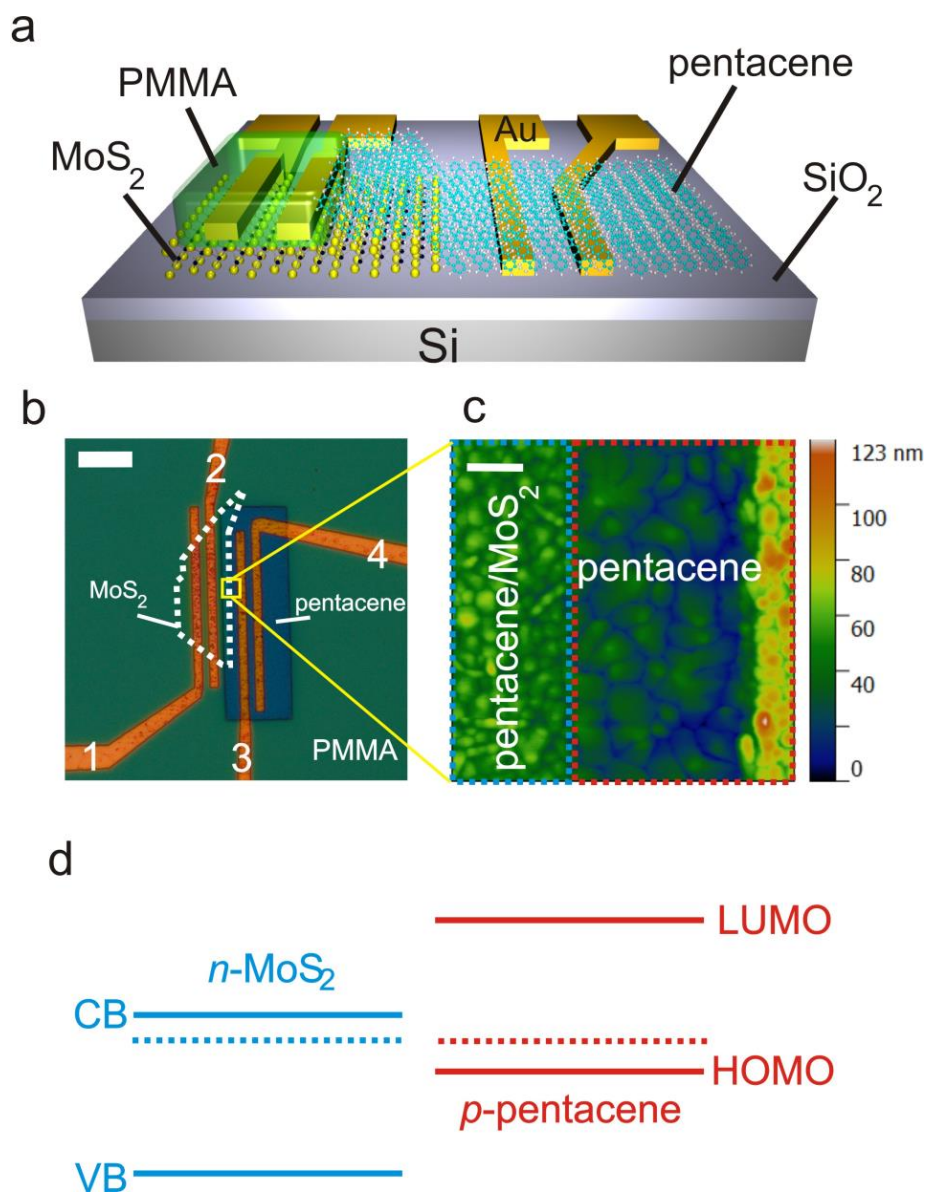

**Figure 1. Structure of the pentacene/MoS₂ p-n heterojunction. a.** Three-dimensional device schematic. Most of the MoS$_2$ flake and electrodes are covered with PMMA to avoid shorting of electrodes by the thermally evaporated pentacene. **b.** Optical micrograph of a representative MoS$_2$/40 nm pentacene device. The MoS$_2$ flake boundary is indicated by the white dashed line. The darker colored rectangle is an opening in the PMMA that exposes the gold pentacene bottom electrodes and a portion of the MoS$_2$ flake (Scale bar = 10 μm). Electrodes 1-2 define an n-type MoS$_2$ FET, 3-4 define a p-type pentacene FET, and 2-3 define the p-n heterojunction device. **c.**

AFM topography image of the junction area indicated by the yellow square in **b.** The pentacene grain size is smaller on $MoS_2$ compared to the surrounding $SiO_2$ substrate. The increased height at the right edge corresponds to the Au electrode below the 40 nm thick pentacene layer (Scale bar = 500 nm). **d.** Band alignments between ultrathin $MoS_2$ and pentacene showing the conduction band (CB) and valence band (VB) for $MoS_2$ and the highest occupied and lowest unoccupied molecular orbital (HOMO and LUMO) levels for pentacene.

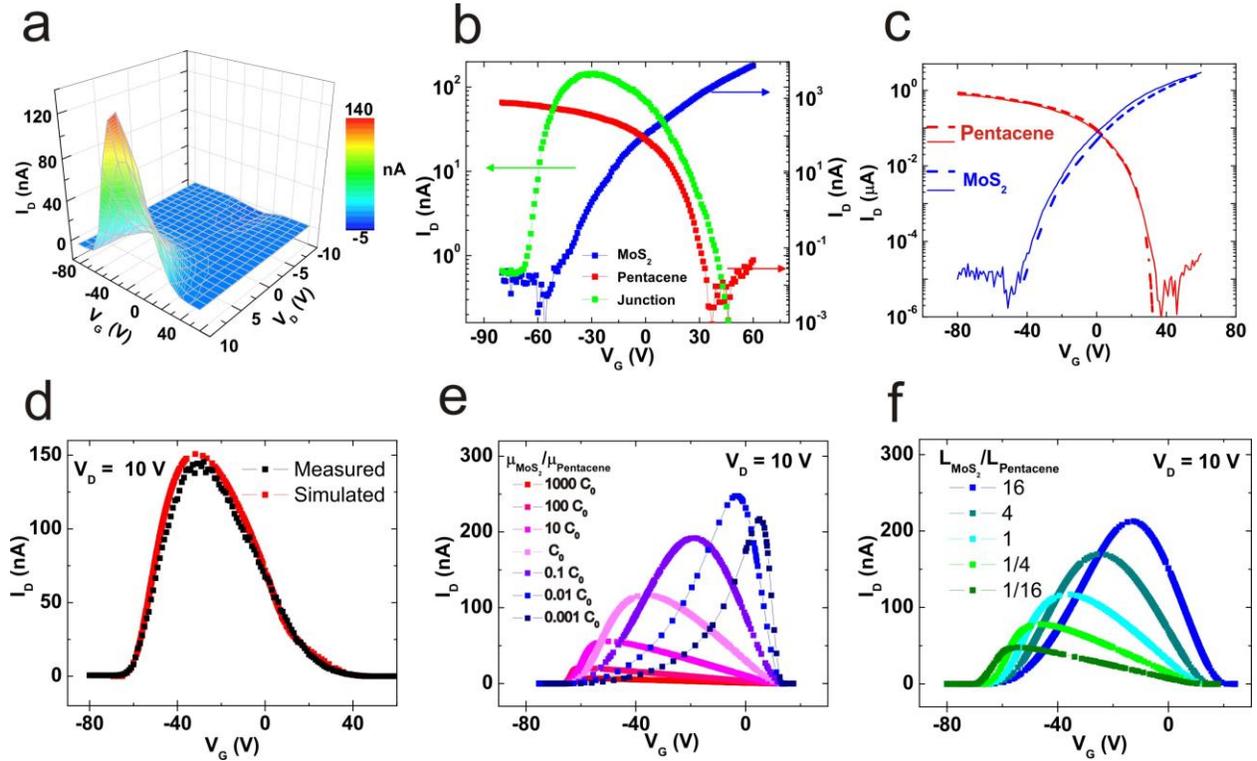

**Figure 2. Electrical properties of the pentacene/MoS$_2$ anti-ambipolar p-n heterojunctions. a.** Gate-dependent $I_D$-$V_D$ (output) characteristics of a representative device. The large current under forward bias and negligible current under reverse bias demonstrate the rectifying nature of the junction. **b.** Semi-log transfer characteristics of the pentacene (red, $V_D = 10$ V) and MoS$_2$ (blue, $V_D = 1$ V) FETs as well as the junction (green, $V_D = 10$ V). The junction transfer curve shows an asymmetric anti-ambipolar characteristic. **c.** Simulated (dashed lines) and experimental (solid lines) transfer characteristics of pentacene (red) and MoS$_2$ (blue) FETs with measured field-effect mobilities of 0.004 and 1.7 cm$^2$/V·s. **d.** Simulated anti-ambipolar transfer plot of the junction (red) along with the measured transfer characteristic (black). Simulated junction transfer characteristics showing the tuning of the anti-ambipolar asymmetry by varying (**e**) ratio of the MoS$_2$ to pentacene mobility values and (**f**) non-overlapped channel lengths. The product of the varied quantity ($\mu$ or $L$) is held constant, where the proportionality constant $C_0$ is the ratio of the mobilities for the device in (**d**).

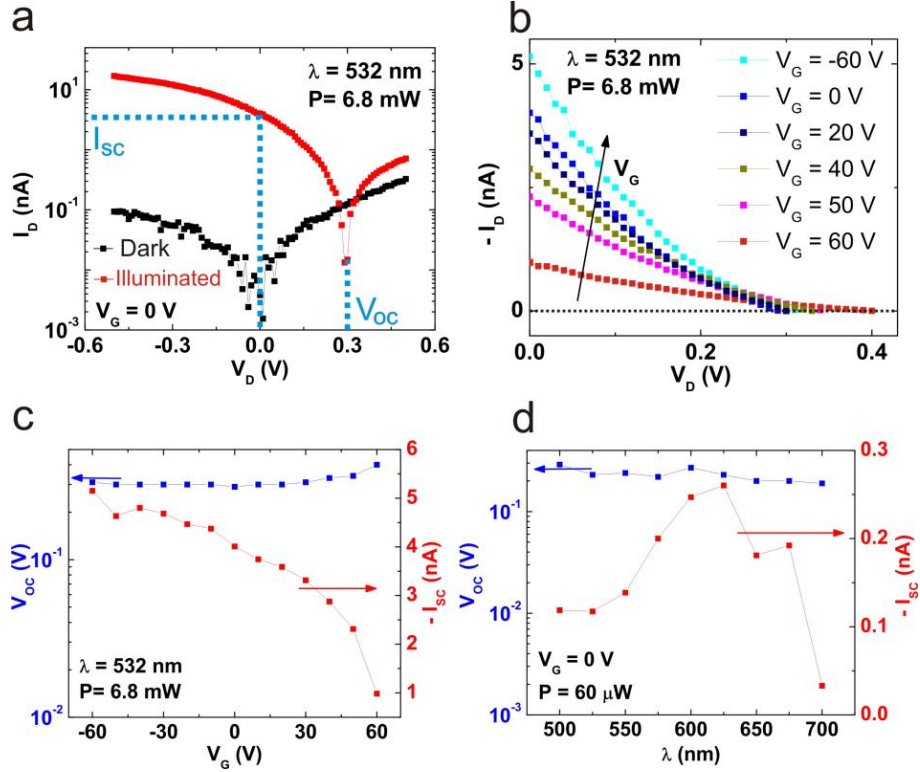

**Figure 3. Photovoltaic effect in pentacene/MoS$_2$ p-n heterojunctions. a.** Dark (black) and illuminated (red) semi-log $I_D$-$V_D$ characteristics of the junction at $V_G = 0$ V (irradiation spot size is 1-2 μm and is centered on the edge of the overlapped pentacene/MoS$_2$ region). A photovoltaic effect with a $V_{OC} = 0.3$ V and $I_{SC} = -3$ nA is observed. **b.** Illuminated $I_D$-$V_D$ characteristics of the junction at varying gate biases. The area under the curve increases with decreasing $V_G$ value. The arrow indicates $V_G$ transitioning from positive to negative values. **c.** Gate dependence of $V_{OC}$ and $I_{SC}$ showing constant $V_{OC}$ while $I_{SC}$ steadily increases with decreasing $V_G$. **d.** Spectral dependence of $V_{OC}$ and $I_{SC}$ as a function of incident wavelength.

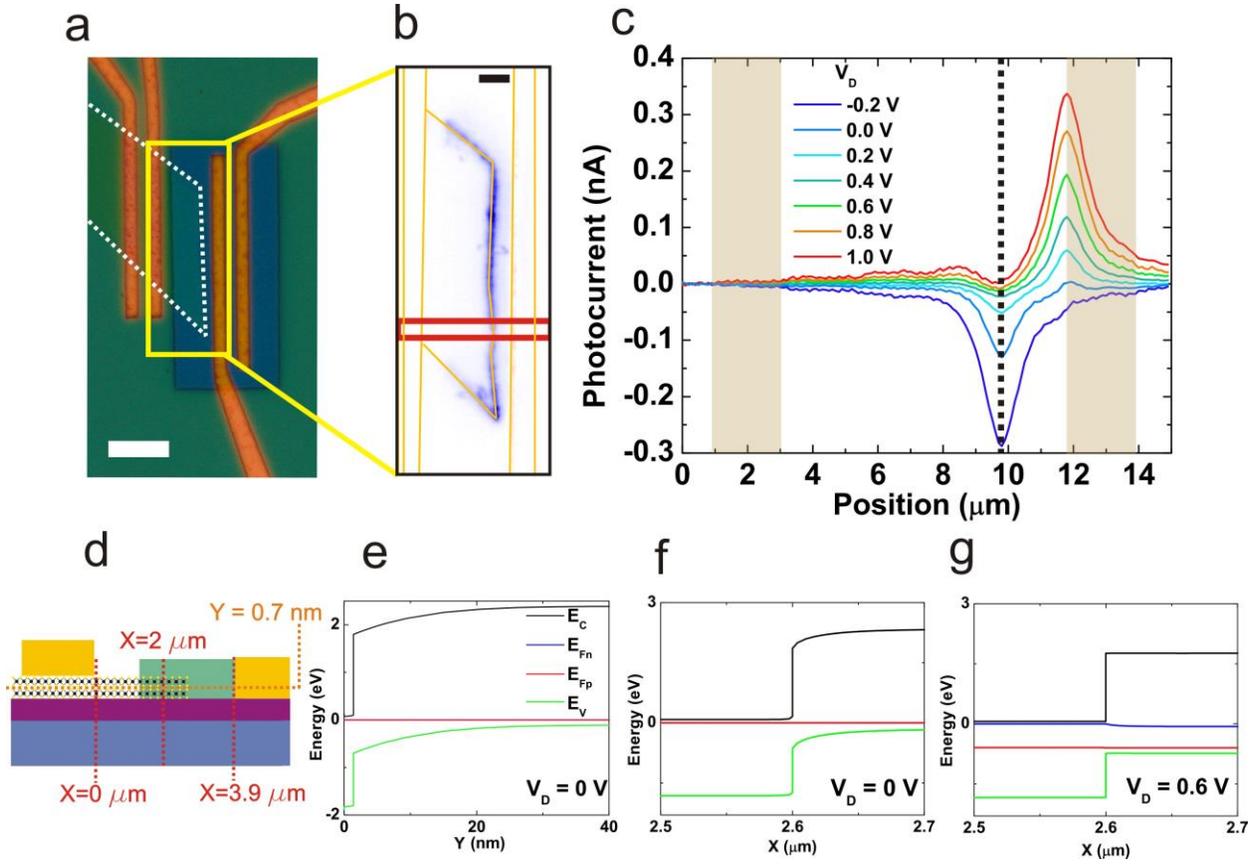

**Figure 4. Spatial and bias dependence of photocurrent. a.** Optical micrograph of a representative device (2L MoS$_2$, 40 nm pentacene) that was scanned under a laser beam (Scale bar = 10 µm). **b.** Spatial photocurrent map at $V_D$ = 0 V of the region indicated by the yellow rectangle in the optical micrograph. The largest photocurrent (blue region) is observed at the edge of the MoS$_2$ flake, while there is negligible photocurrent response from the contacts on either side of the junction (Scale bar = 3 µm) **c.** Photocurrent line profiles averaged over the red rectangle shown in (**b**) at varying $V_D$ values. The gold rectangles indicate the position of the electrodes while the black dashed line indicates the end of the overlapped pentacene/MoS$_2$ region. **d.** Schematic diagram of the simulated junction device geometry. **e.** Band profiles at $V_D$ = 0 V across a vertical cross section of the junction at X = 2 µm indicating the presence of an electric field normal to the interface. **f.** Horizontal band profiles near the edge of the overlapped pentacene/MoS$_2$ region (X = 2.6 µm). **g.** Band alignments in a horizontal cross-section at the flat band condition ($V_D$ = 0.6 V). E$_c$ and E$_v$ represent conduction and valence band energies, respectively, while E$_{Fn}$ and E$_{Fp}$ represent quasi-Fermi levels for electrons and holes, respectively.

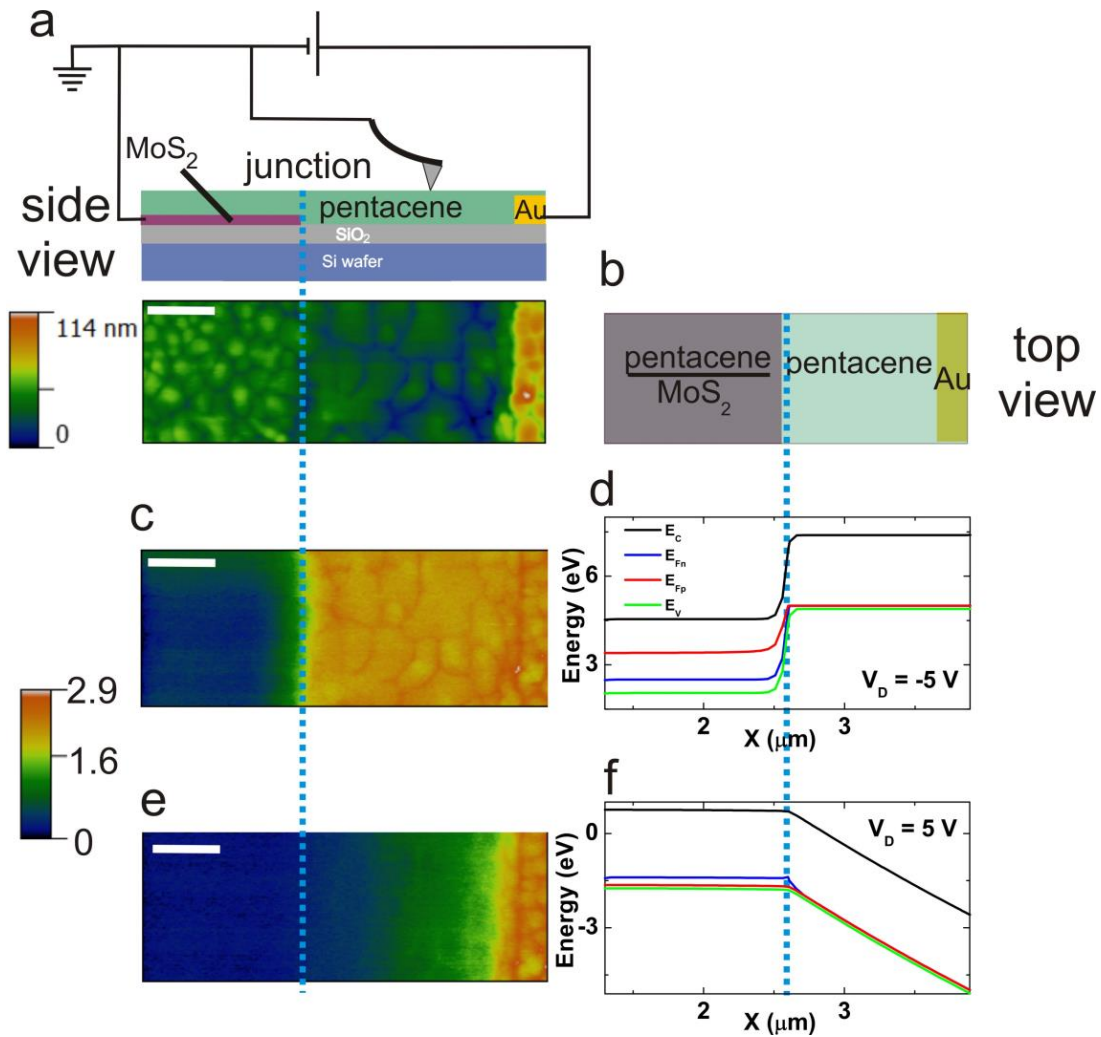

**Figure 5. Electrostatic force microscopy. a**. Schematic side view of the measurement setup and device (top) with an AFM topography image of the device in Figure 1 with 2L MoS$_2$/40 nm pentacene (bottom). **b.** Top view of the device with the labelled regions. **c.** EFM phase map of the region shown in (**a**) at $V_D$ = -5 V. **d.** Corresponding simulated band diagram at the surface showing the potential drop at the edge of the overlapped pentacene/MoS$_2$ region. **e.** EFM phase map of the region shown in (**a**) at $V_D$ = 5 V. **f.** Corresponding simulated band profiles at the surface showing a gradual potential drop across the pentacene channel (Scale bars = 500 nm). The dashed blue line indicates the edge of the overlapped pentacene/MoS$_2$ region.

# Supporting Information

# Hybrid, Gate-Tunable, van der Waals p-n Heterojunctions from Pentacene and MoS$_2$

*Deep Jariwala[1†], Sarah L. Howell[1†], Kan-Sheng Chen[1], Junmo Kang[1], Vinod K. Sangwan[1], Stephen A. Filippone[1], Riccardo Turrisi[2], Tobin J. Marks[1,2]\*, Lincoln J. Lauhon[1]\* and Mark C. Hersam[1,2]\**

[1]Department of Materials Science and Engineering, Northwestern University, Evanston, Illinois 60208, USA.

[2]Department of Chemistry, Northwestern University, Evanston, Illinois 60208, USA.

† These authors contributed equally.

*e-mail: m-hersam@northwestern.edu, lauhon@northwestern.edu, t-marks@northwestern.edu

**S1. Material synthesis, deposition, fabrication and characterization**

MoS$_2$ crystals were purchased from SPI Supplies and mechanically exfoliated using Scotch tape on thermally oxidized 300 nm SiO$_2$/Si wafers. The Si wafers (Si orientation <100>) were purchased from Silicon Quest International. The wafers were doped degenerately n-type with As (resistivity = 0.001-0.005 Ω-cm). The MoS$_2$ flakes were exfoliated on the wafers after solvent cleaning and sonication in acetone and isopropanol (IPA). Raman spectra were acquired using a confocal Raman microscope (Horiba Xplora Plus) using a 532 nm excitation line and a 1800 lines/mm grating with a 100x objective and an acquisition time of 30 sec. Fig. S1 shows Raman spectra of representative monolayer and bilayer flakes.

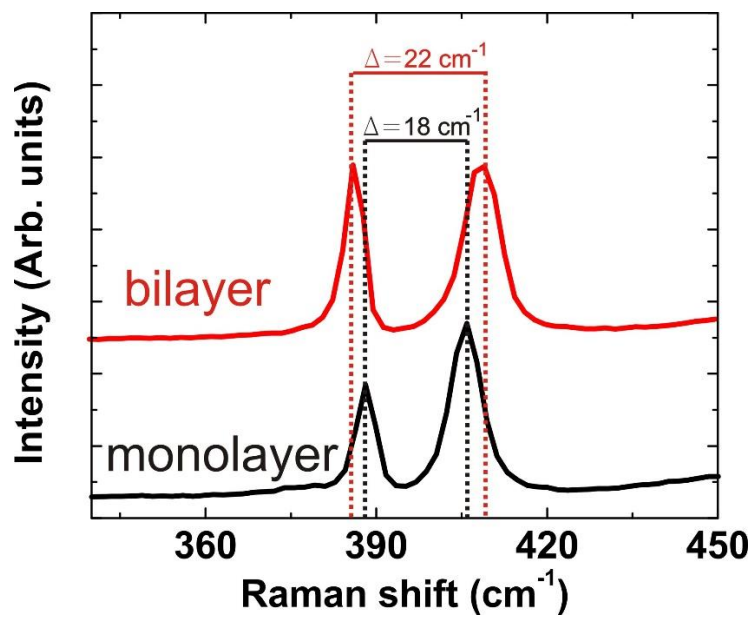

**Figure S1.** Raman spectra of representative monolayer and bilayer $MoS_2$ flakes used in the present study. The separation between $E_{2g}$ (left) and $A_{1g}$ (right) peaks for the black curve is < 20 $cm^{-1}$, which is a signature of monolayer $MoS_2$, while a separation of 22 $cm^{-1}$ (red curve) indicates bilayer $MoS_2$ as per prior literature reports.[1] The spectra are offset along the y axis for visual clarity.

Electron beam lithography was used for fabrication with PMMA A4 (Microchem) as the resist, which was coated at a spin speed of 4000 rpm for 45 s and baked on a hot-plate at 180 °C for 90 s. An electron dose of 360 $\mu C/cm^2$ was used for exposure followed by development in a methyl iso butyl ketone (MIBK):IPA (1:3) solution for 75 s. The contact metals were evaporated using a thermal evaporator (Lesker Nano 38). A stack of 3.5 nm thick Ti and 34.5 nm thick Au was used to make a bottom contact of Au with pentacene and a top contact of Ti with $MoS_2$. Following contact metal deposition, a second step of lithography was performed to open a window in the PMMA for deposition of pentacene. This window also exposed part of the $MoS_2$ flake for the pentacene/$MoS_2$ heterojunction. A shadow mask (1 mm x 1 mm) was placed such that the PMMA window lies inside the mask but the metal pads for probing are outside the mask.

Prior to use, pentacene (99%, Sigma-Aldrich) was sublimed twice (base pressure: $7.0 \times 10^{-6}$ Torr) in a three-zone thermal gradient sublimator connected to a diffusion pump. The sublimation temperature was kept below 300°C to avoid decomposition of pentacene.[2] Zone temperatures: 290°C, 275°C, 230°C. Only material from Zone 3 (230°C) was collected and used for the experiments reported herein. The purified powder (dark violet color) was then loaded into a tungsten boat and evaporated in a thermal evaporator. The evaporation rate was maintained constant at 0.15-0.2 Å/s, and the substrate was held at room temperature. The thickness of the pentacene film was between 40 and 45 nm. An absorption spectrum of a ~40 nm thick pentacene film acquired using a Cary 5000 spectrophotometer (Agilent Technologies) is shown in Fig. S2.

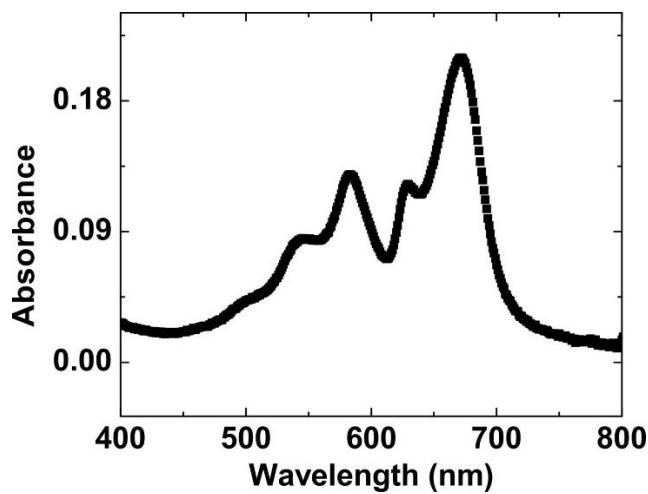

**Figure S2.** Absorption spectrum of a thermally evaporated 40 nm thick pentacene film on a glass slide. The film was deposited during the same deposition as one of the heterojunction devices presented in the manuscript. The observed spectrum agrees well with prior literature reports.[3,4]

**S2. Electrical and photocurrent measurements**

All electrical measurements were performed under high vacuum ($P < 5 \times 10^{-5}$ Torr) in a Lake Shore CRX 4K probe station. Output and transfer characteristics were measured using

Keithley 2400 source meters and custom LabView programs. The gate voltage was swept at 10 V/sec in steps of 1 V in the transfer plots. The asymmetry in transconductances observed in the anti-ambipolar transfer characteristics of the junction are quantified by taking the ratio of the transconductance on either side of the peak. This ratio is ~2.1 in the device shown in Fig. 2. We observe device-to-device variations in this ratio which ranges from 1.2 to 2.4. Experimentally, these variations arise due to differences in size and electrical properties (e.g., mobility) of exfoliated $MoS_2$ flakes.

The photovoltaic (PV) measurements were performed by wire bonding the device inside an optical cryostat with a transparent window. The cryostat was pumped down using a roughing pump. A CW laser was used for illumination at 532 nm, while a tunable coherent white light source (NKT Photonics) was used for other wavelengths.

The scanning photocurrent measurements were performed using a scanning confocal microscope (WiTec) in air. The tunable coherent white light source (NKT Photonics) was used to generate the spatially and spectrally resolved photocurrent. The photocurrent is converted into a voltage by a current preamplifier and recorded by a lock-in amplifier (for imaging). The reflected light is also recorded simultaneously to correlate the spatial photocurrent map with the device geometry.

## S3. Surface potential and topography measurements

AFM topography and electrostatic force microscopy (EFM) were performed using an Asylum Cypher S system. The topography images were taken with tapping mode in the repulsive regime. For EFM characterization, Pt-Ir coated conductive AFM tips with resonance frequencies of ~75 kHz were used. During EFM measurements, the cantilever amplitude was ~30 nm when acquiring the topographic profile of the sample and ~3 nm when acquiring the electrostatic force profile. When acquiring the electrostatic force profile, the cantilever was maintained 50 nm above the sample surface using the topographic profile obtained in tapping mode.

## S4. Finite element simulations

Sentaurus TCAD was used to model $MoS_2$ devices by solving the following steady-state coupled coupled differential equations in two dimensions:

| | |
|---|---|
| $\nabla \cdot (\varepsilon \nabla \varphi) = -q(p - n + N_D - N_A) - \rho_{trap}$ | Poisson's equation |
| $\nabla \cdot J_n = q^* R_{net}$ | Electron continuity equation |
| $-\nabla \cdot J_p = q^* R_{net}$ | Hole continuity equation |
| $J_n = -n \cdot q \cdot \mu_n (\nabla \Phi_n)$ | Electron current equation |
| $J_p = -p \cdot q \cdot \mu_p (\nabla \Phi_p)$ | Hole current equation |

where $\varepsilon$ is the electrical permittivity, $\varphi$ is the electrostatic potential, q is the charge of an electron, p and n are the hole and electron densities, $N_D$ and $N_A$ are the donor and acceptor concentrations, $\rho_{trap}$ is the charge density contributed by traps, $J_n$ and $J_p$ are the electron and hole current densities, $R_{net}$ is the net recombination rate, $\mu_n$ and $\mu_p$ are the electron and hole mobilities, and $\Phi_n$ and $\Phi_p$ are the electron and hole quasi-Fermi potentials.

The simulated geometry is shown in Fig. 4d of the manuscript where the 1.4 nm thick $MoS_2$ channel, the pentacene/$MoS_2$ overlap region, and the 40 nm thick pentacene channel were each 1.3 um long and on a 300 nm $SiO_2$ layer on the conductive substrate, which serves as the back gate. When available, the materials parameters used in the simulations were taken from the literature. Otherwise material parameters were determined by experimental measurements on uniform FETs for the respective channel materials. The model assumes a three-dimensional effective density of states, Fermi-Dirac carrier statistics, and complete ionization of dopants. The contacts are modeled as Schottky-type, and transport though the semiconductor heterojunction is governed by a thermionic emission model. No interband tunneling contributions were included in the modeling of currents under forward bias since we expect interband tunneling will only play a significant role in charge transport under reverse bias. A carrier concentration dependent mobility model was used for MoS$_2$[5] and a field dependent mobility model was used for pentacene[6]:

$$\mu_{MoS2} = \frac{3500}{N_I/10^{11}cm^{-2}} \left[A(\varepsilon_e) + (\frac{n_s}{10^{13}cm^{-2}})^{1.2}\right] cm^2/Vs$$

$$\mu_{pentacene} = \mu_0 e^{\sqrt{E/E_0}}$$

Materials parameters for MoS$_2$ and pentacene were identified such that the simulated transfer curves reproduced those measured in Fig. 2b for individual FETs devices, and the corresponding material properties are listed in Table S1. Only the mobility multiplier (μ$_0$ and N$_I$) and the trap concentrations in MoS$_2$ and pentacene were used as fitting parameters for Figure 2c. Since the MoS$_2$ FET was measured at V$_D$ = 0.5 V while the junction was measured under the high bias of V$_D$ = 10 V, we account for the fact that the electron mobility in ultrathin MoS$_2$ depends on applied lateral electric fields[7-9] by decreasing the MoS$_2$ mobility multiplier used for the junction device simulations. Only the pentacene mobility multiplier (μ$_0$) and the trap concentrations in MoS$_2$ and pentacene were used as fitting parameters for the anti-ambipolar transfer characteristic in Figure 2d, and the corresponding material properties are listed in Table S2. The nominal variations in mobility and trap concentrations between Table S1 and S2 are expected since the dielectric environment experienced by both semiconductors in the overlapped region of the junction device (where the two materials are in intimate contact) is different compared to control FETs where they are isolated. For example, the pentacene material grown on MoS$_2$ has smaller grain boundaries, which leads to additional grain boundary scattering and trapping, and device-to-device variations.

**Table S1.** Material constants used in the simulations of control FETs in Figure 2c.

| MoS$_2$ parameters | Values |
|---|---|
| Thickness | 2L, 1.4 nm |
| Doping density | 1.5e18 cm$^{-3}$ |
| Affinity, Band Gap | 4.2 eV, 1.9 eV |
| Mobility model parameters | $N_I$=3.8e21 cm$^{-3}$, $A(\varepsilon_e)$=0.05 |
| Traps 0.1eV below conduction band | 12e18 cm$^{-3}$ |
| Traps 0.2eV below conduction band | 6e18 cm$^{-3}$ |
| Traps 0.3eV below conduction band | 6e18 cm$^{-3}$ |
| Ti contact affinity | 4.3 eV |
| **Pentacene parameters** | **Values** |
| Thickness | 40 nm |
| Doping density | -6.6e16 cm$^{-3}$ |
| Affinity, Band Gap | 2.5 eV, 2.5 eV |
| Mobility model parameters | $\mu_0$=2.8e-3 cm$^2$/V/s, $E_0$=3e5 V/cm |
| Nv | 1e19 cm$^{-3}$ |
| Dielectric constant | 4 |
| Traps 0.1eV below valence band | 4.2e16 cm$^{-3}$ |
| Traps 0.2eV below valence band | 16.8e16 cm$^{-3}$ |
| Traps 0.3eV below valence band | 12.6e16 cm$^{-3}$ |
| Au contact affinity | 4.9 |

**Table S2.** Material constants used in the simulations of junctions in Figure 2d-e, Figure 4e-g and Figure 5d,f.

| MoS$_2$ parameters | values |
|---|---|
| Thickness | 2L, 1.4 nm |
| Doping density | 1.5e18 cm$^{-3}$ |
| Affinity, Band Gap | 4.2 eV, 1.9 eV |
| Mobility model parameters | $N_I$=6.9e20 cm$^{-3}$, $A(\varepsilon_e)$=0.05 |
| Traps 0.1eV below conduction band | 28e18 cm$^{-3}$ |
| Traps 0.2eV below conduction band | 4e18 cm$^{-3}$ |
| Traps 0.3eV below conduction band | 0e18 cm$^{-3}$ |
| Ti contact affinity | 4.3 eV |
| **Pentacene parameters** | **values** |
| Thickness | 40 nm |
| Doping density | -6.6e16 cm$^{-3}$ |
| Affinity, Band Gap | 2.5 eV, 2.5 eV |
| Mobility model parameters | $\mu_0$=1.9e-3 cm$^2$/V/s, $E_0$=3e5 V/cm |
| Nv | 1e19 cm$^{-3}$ |
| Dielectric constant | 4 |
| Traps 0.0eV below valence band | 9e16 cm$^{-3}$ |
| Traps 0.2eV below valence band | 34e16 cm$^{-3}$ |
| Traps 0.4eV below valence band | 25e16 cm$^{-3}$ |
| Au contact affinity | 4.9 |